

\magnification=\magstephalf


\newbox\SlashedBox
\def\slashed#1{\setbox\SlashedBox=\hbox{#1}
\hbox to 0pt{\hbox to 1\wd\SlashedBox{\hfil/\hfil}\hss}#1}
\def\hboxtosizeof#1#2{\setbox\SlashedBox=\hbox{#1}
\hbox to 1\wd\SlashedBox{#2}}

\def\mathslashed#1{\setbox\SlashedBox=\hbox{$#1$}
\hbox to 0pt{\hbox to 1\wd\SlashedBox{\hfil/\hfil}\hss}#1}

\def\ifsmall{\iffalse}  
\def\titlepagefont{}  

\def\DefineTeXgraphics{%
\special{ps::[global] /TeXgraphics { } def}}  

\def\today{\ifcase\month\or January\or February\or March\or April\or May
\or June\or July\or August\or September\or October\or November\or
December\fi\space\number\day, \number\year}
\def\eatPrefix19{}
\def\Year{\expandafter\eatPrefix\the\year}
\newcount\hours \newcount\minutes
\def\monthname{\ifcase\month\or
January\or February\or March\or April\or May\or June\or July\or
August\or September\or October\or November\or December\fi}
\def\shortmonthname{\ifcase\month\or
Jan\or Feb\or Mar\or Apr\or May\or Jun\or Jul\or
Aug\or Sep\or Oct\or Nov\or Dec\fi}

\def\TimeStamp{\hours\the\time\divide\hours by60%
\minutes -\the\time\divide\minutes by60\multiply\minutes by60%
\advance\minutes by\the\time%
${\rm \shortmonthname}\cdot\if\day<10{}0\fi\the\day\cdot\the\year%
\qquad\the\hours:\if\minutes<10{}0\fi\the\minutes$}




\def\Title#1{%
\vskip 1in{\titlefont\centerline{#1}}\vskip .5in}

\def\Date#1{\leftline{#1}\tenrm\supereject%
\global\hsize=\hsbody\global\hoffset=\hbodyoffset%
\footline={\hss\tenrm\folio\hss}}

\newif\ifdraftmode
\newif\ifleftlabels  

\def\nolabels{\def\wrlabeL##1{}\def\eqlabeL##1{}\def\reflabeL##1{}}
\def\writelabels{\def\wrlabeL##1{\leavevmode\vadjust{\rlap{\smash%
{\line{{\escapechar=` \hfill\rlap{\sevenrm\hskip.03in\string##1}}}}}}}%
\def\eqlabeL##1{{\escapechar-1\rlap{\sevenrm\hskip.05in\string##1}}}%
\def\reflabeL##1{\noexpand\rlap{\noexpand\sevenrm[\string##1]}}}
\def\writeleftlabels{\def\wrlabeL##1{\leavevmode\vadjust{\rlap{\smash%
{\line{{\escapechar=` \hfill\rlap{\sevenrm\hskip.03in\string##1}}}}}}}%
\def\eqlabeL##1{{\escapechar-1%
\rlap{\sixrm\hskip.05in\string##1}%
\llap{\sevenrm\string##1\hskip.03in\hbox to \hsize{}}}}%
\def\reflabeL##1{\noexpand\rlap{\noexpand\sevenrm[\string##1]}}}
\nolabels

\newdimen\fullhsize
\newdimen\hstitle
\hstitle=\hsize 
\newdimen\hsbody
\hsbody=\hsize 
\newdimen\hbodyoffset
\hbodyoffset=\hoffset 
\newbox\leftpage
\def\abstract#1{#1}
\def\rotated{\special{ps: landscape}
\magnification=1000  
\baselineskip=14pt
\global\hstitle=9truein\global\hsbody=4.75truein
\global\vsize=7truein\global\voffset=-.31truein
\global\hoffset=-0.54in\global\hbodyoffset=-.54truein
\global\fullhsize=10truein
\def\DefineTeXgraphics{%
\special{ps::[global]
/TeXgraphics {currentpoint translate 0.7 0.7 scale
              -80 0.72 mul -1000 0.72 mul translate} def}}
\let\lr=L
\def\ifsmall{\iftrue}
\def\titlepagefont{\twelvepoint}
\trueseventeenpoint
\def\almostshipout##1{\if L\lr \count1=1
      \global\setbox\leftpage=##1 \global\let\lr=R
   \else \count1=2
      \shipout\vbox{\hbox to\fullhsize{\box\leftpage\hfil##1}}
      \global\let\lr=L\fi}

\output={\ifnum\count0=1 
 \shipout\vbox{\hbox to \fullhsize{\hfill\pagebody\hfill}}\advancepageno
 \else
 \almostshipout{\leftline{\vbox{\pagebody\makefootline}}}\advancepageno
 \fi}

\def\abstract##1{{\leftskip=1.5in\rightskip=1.5in ##1\par}} }

\def\linemessage#1{\immediate\write16{#1}}

\global\newcount\secno \global\secno=0
\global\newcount\appno \global\appno=0
\global\newcount\meqno \global\meqno=1
\global\newcount\subsecno \global\subsecno=0
\global\newcount\figno \global\figno=0

\newif\ifAnyCounterChanged
\let\terminator=\relax
\def\normalize#1{\ifx#1\terminator\let\next=\relax\else%
\if#1i\aftergroup i\else\if#1v\aftergroup v\else\if#1x\aftergroup x%
\else\if#1l\aftergroup l\else\if#1c\aftergroup c\else%
\if#1m\aftergroup m\else%
\if#1I\aftergroup I\else\if#1V\aftergroup V\else\if#1X\aftergroup X%
\else\if#1L\aftergroup L\else\if#1C\aftergroup C\else%
\if#1M\aftergroup M\else\aftergroup#1\fi\fi\fi\fi\fi\fi\fi\fi\fi\fi\fi\fi%
\let\next=\normalize\fi%
\next}
\def\makeNormal#1#2{\def\doNormalDef{\edef#1}\begingroup%
\aftergroup\doNormalDef\aftergroup{\normalize#2\terminator\aftergroup}%
\endgroup}

\def\warnIfChanged#1#2{%
\ifundef#1
\else\begingroup%
\edef\oldDefinitionOfCounter{#1}\edef\newDefinitionOfCounter{#2}%
\ifx\oldDefinitionOfCounter\newDefinitionOfCounter%
\else%
\linemessage{Warning: definition of \noexpand#1 has changed.}%
\global\AnyCounterChangedtrue\fi\endgroup\fi}

\def\Section#1{\global\advance\secno by1\relax\global\meqno=1%
\global\subsecno=0%
\bigbreak\bigskip
\centerline{\twelvepoint \bf %
\the\secno. #1}%
\par\nobreak\medskip\nobreak}
\def\tagsection#1{%
\warnIfChanged#1{\the\secno}%
\xdef#1{\the\secno}%
\ifWritingAuxFile\immediate\write\auxfile{\noexpand\xdef\noexpand#1{#1}}\fi%
}
\def\section{\Section}
\def\Subsection#1{\global\advance\subsecno by1\relax\medskip %
\leftline{\bf\the\secno.\the\subsecno\ #1}%
\par\nobreak\smallskip\nobreak}
\def\tagsubsection#1{%
\warnIfChanged#1{\the\secno.\the\subsecno}%
\xdef#1{\the\secno.\the\subsecno}%
\ifWritingAuxFile\immediate\write\auxfile{\noexpand\xdef\noexpand#1{#1}}\fi%
}

\def\subsection{\Subsection}

\def\romappno{\uppercase\expandafter{\romannumeral\appno}}
\def\makeNormalizedRomappno{%
\expandafter\makeNormal\expandafter\normalizedromappno%
\expandafter{\romannumeral\appno}%
\edef\normalizedromappno{\uppercase{\normalizedromappno}}}
\def\Appendix#1{\global\advance\appno by1\relax\global\meqno=1\global\secno=0
\bigbreak\bigskip
\centerline{\twelvepoint \bf Appendix %
\romappno. #1}%
\par\nobreak\medskip\nobreak}
\def\tagappendix#1{\makeNormalizedRomappno%
\warnIfChanged#1{\normalizedromappno}%
\xdef#1{\normalizedromappno}%
\ifWritingAuxFile\immediate\write\auxfile{\noexpand\xdef\noexpand#1{#1}}\fi%
}
\def\appendix{\Appendix}

\def\eqn#1{\makeNormalizedRomappno%
\ifnum\secno>0%
  \warnIfChanged#1{\the\secno.\the\meqno}%
  \eqno(\the\secno.\the\meqno)\xdef#1{\the\secno.\the\meqno}%
     \global\advance\meqno by1
\else\ifnum\appno>0%
  \warnIfChanged#1{\normalizedromappno.\the\meqno}%
  \eqno({\rm\romappno}.\the\meqno)%
      \xdef#1{\normalizedromappno.\the\meqno}%
     \global\advance\meqno by1
\else%
  \warnIfChanged#1{\the\meqno}%
  \eqno(\the\meqno)\xdef#1{\the\meqno}%
     \global\advance\meqno by1
\fi\fi%
\eqlabeL#1%
\ifWritingAuxFile\immediate\write\auxfile{\noexpand\xdef\noexpand#1{#1}}\fi%
}
\def\defeqn#1{\makeNormalizedRomappno%
\ifnum\secno>0%
  \warnIfChanged#1{\the\secno.\the\meqno}%
  \xdef#1{\the\secno.\the\meqno}%
     \global\advance\meqno by1
\else\ifnum\appno>0%
  \warnIfChanged#1{\normalizedromappno.\the\meqno}%
  \xdef#1{\normalizedromappno.\the\meqno}%
     \global\advance\meqno by1
\else%
  \warnIfChanged#1{\the\meqno}%
  \xdef#1{\the\meqno}%
     \global\advance\meqno by1
\fi\fi%
\eqlabeL#1%
\ifWritingAuxFile\immediate\write\auxfile{\noexpand\xdef\noexpand#1{#1}}\fi%
}
\def\anoneqn{\makeNormalizedRomappno%
\ifnum\secno>0
  \eqno(\the\secno.\the\meqno)%
     \global\advance\meqno by1
\else\ifnum\appno>0
  \eqno({\rm\normalizedromappno}.\the\meqno)%
     \global\advance\meqno by1
\else
  \eqno(\the\meqno)%
     \global\advance\meqno by1
\fi\fi%
}
\def\mfig#1#2{\global\advance\figno by1%
\relax#1\the\figno%
\warnIfChanged#2{\the\figno}%
\edef#2{\the\figno}%
\reflabeL#2%
\ifWritingAuxFile\immediate\write\auxfile{\noexpand\xdef\noexpand#2{#2}}\fi%
}

\catcode`@=11 

\font\ninerm=cmr9
\font\eightrm=cmr8
\font\sixrm=cmr6

\def\loadtrueseventeenpoint{
 \font\seventeenrm=cmr10 at 17.28truept
 \font\seventeeni=cmmi10 at 17.28truept
 \font\seventeenbf=cmbx10 at 17.28truept
 \font\seventeenit=cmti10 at 17.28truept
 \font\seventeensl=cmsl10 at 17.28truept
 \font\seventeensy=cmsy10 at 17.28truept
}
\def\loadfourteenpoint{
\font\fourteenrm=cmr10 at 14.4pt
\font\fourteeni=cmmi10 at 14.4pt
\font\fourteenit=cmti10 at 14.4pt
\font\fourteensl=cmsl10 at 14.4pt
\font\fourteensy=cmsy10 at 14.4pt
\font\fourteenbf=cmbx10 at 14.4pt
}
\def\loadtruetwelvepoint{
\font\twelverm=cmr10 at 12truept
\font\twelvei=cmmi10 at 12truept
\font\twelveit=cmti10 at 12truept
\font\twelvesl=cmsl10 at 12truept
\font\twelvesy=cmsy10 at 12truept
\font\twelvebf=cmbx10 at 12truept
}

\font\ninei=cmmi9
\font\eighti=cmmi8
\font\sixi=cmmi6
\skewchar\ninei='177 \skewchar\eighti='177 \skewchar\sixi='177

\font\ninesy=cmsy9
\font\eightsy=cmsy8
\font\sixsy=cmsy6
\skewchar\ninesy='60 \skewchar\eightsy='60 \skewchar\sixsy='60

\font\ninebf=cmbx9
\font\eightbf=cmbx8
\font\sixbf=cmbx6

\font\ninett=cmtt9
\font\eighttt=cmtt8

\hyphenchar\tentt=-1 
\hyphenchar\ninett=-1
\hyphenchar\eighttt=-1

\font\ninesl=cmsl9
\font\eightsl=cmsl8

\font\nineit=cmti9
\font\eightit=cmti8


\newskip\ttglue
\def\tenpoint{\def\rm{\fam0\tenrm}%
  \textfont0=\tenrm \scriptfont0=\sevenrm \scriptscriptfont0=\fiverm
  \textfont1=\teni \scriptfont1=\seveni \scriptscriptfont1=\fivei
  \textfont2=\tensy \scriptfont2=\sevensy \scriptscriptfont2=\fivesy
  \textfont3=\tenex \scriptfont3=\tenex \scriptscriptfont3=\tenex
  \def\it{\fam\itfam\tenit}\textfont\itfam=\tenit
  \def\sl{\fam\slfam\tensl}\textfont\slfam=\tensl
  \def\bf{\fam\bffam\tenbf}\textfont\bffam=\tenbf \scriptfont\bffam=\sevenbf
  \scriptscriptfont\bffam=\fivebf
  \normalbaselineskip=12pt
  \let\sc=\eightrm
  \let\big=\tenbig
  \setbox\strutbox=\hbox{\vrule height8.5pt depth3.5pt width\z@}%
  \normalbaselines\rm}

\def\twelvepoint{\def\rm{\fam0\twelverm}%
  \textfont0=\twelverm \scriptfont0=\ninerm \scriptscriptfont0=\sevenrm
  \textfont1=\twelvei \scriptfont1=\ninei \scriptscriptfont1=\seveni
  \textfont2=\twelvesy \scriptfont2=\ninesy \scriptscriptfont2=\sevensy
  \textfont3=\tenex \scriptfont3=\tenex \scriptscriptfont3=\tenex
  \def\it{\fam\itfam\twelveit}\textfont\itfam=\twelveit
  \def\sl{\fam\slfam\twelvesl}\textfont\slfam=\twelvesl
  \def\bf{\fam\bffam\twelvebf}\textfont\bffam=\twelvebf
\scriptfont\bffam=\ninebf
  \scriptscriptfont\bffam=\sevenbf
  \normalbaselineskip=12pt
  \let\sc=\eightrm
  \let\big=\tenbig
  \setbox\strutbox=\hbox{\vrule height8.5pt depth3.5pt width\z@}%
  \normalbaselines\rm}

\def\fourteenpoint{\def\rm{\fam0\fourteenrm}%
  \textfont0=\fourteenrm \scriptfont0=\tenrm \scriptscriptfont0=\sevenrm
  \textfont1=\fourteeni \scriptfont1=\teni \scriptscriptfont1=\seveni
  \textfont2=\fourteensy \scriptfont2=\tensy \scriptscriptfont2=\sevensy
  \textfont3=\tenex \scriptfont3=\tenex \scriptscriptfont3=\tenex
  \def\it{\fam\itfam\fourteenit}\textfont\itfam=\fourteenit
  \def\sl{\fam\slfam\fourteensl}\textfont\slfam=\fourteensl
  \def\bf{\fam\bffam\fourteenbf}\textfont\bffam=\fourteenbf%
  \scriptfont\bffam=\tenbf
  \scriptscriptfont\bffam=\sevenbf
  \normalbaselineskip=17pt
  \let\sc=\elevenrm
  \let\big=\tenbig
  \setbox\strutbox=\hbox{\vrule height8.5pt depth3.5pt width\z@}%
  \normalbaselines\rm}

\def\seventeenpoint{\def\rm{\fam0\seventeenrm}%
  \textfont0=\seventeenrm \scriptfont0=\fourteenrm \scriptscriptfont0=\tenrm
  \textfont1=\seventeeni \scriptfont1=\fourteeni \scriptscriptfont1=\teni
  \textfont2=\seventeensy \scriptfont2=\fourteensy \scriptscriptfont2=\tensy
  \textfont3=\tenex \scriptfont3=\tenex \scriptscriptfont3=\tenex
  \def\it{\fam\itfam\seventeenit}\textfont\itfam=\seventeenit
  \def\sl{\fam\slfam\seventeensl}\textfont\slfam=\seventeensl
  \def\bf{\fam\bffam\seventeenbf}\textfont\bffam=\seventeenbf%
  \scriptfont\bffam=\fourteenbf
  \scriptscriptfont\bffam=\twelvebf
  \normalbaselineskip=21pt
  \let\sc=\fourteenrm
  \let\big=\tenbig
  \setbox\strutbox=\hbox{\vrule height 12pt depth 6pt width\z@}%
  \normalbaselines\rm}

\def\ninepoint{\def\rm{\fam0\ninerm}%
  \textfont0=\ninerm \scriptfont0=\sixrm \scriptscriptfont0=\fiverm
  \textfont1=\ninei \scriptfont1=\sixi \scriptscriptfont1=\fivei
  \textfont2=\ninesy \scriptfont2=\sixsy \scriptscriptfont2=\fivesy
  \textfont3=\tenex \scriptfont3=\tenex \scriptscriptfont3=\tenex
  \def\it{\fam\itfam\nineit}\textfont\itfam=\nineit
  \def\sl{\fam\slfam\ninesl}\textfont\slfam=\ninesl
  \def\bf{\fam\bffam\ninebf}\textfont\bffam=\ninebf \scriptfont\bffam=\sixbf
  \scriptscriptfont\bffam=\fivebf
  \normalbaselineskip=11pt
  \let\sc=\sevenrm
  \let\big=\ninebig
  \setbox\strutbox=\hbox{\vrule height8pt depth3pt width\z@}%
  \normalbaselines\rm}

\def\eightpoint{\def\rm{\fam0\eightrm}%
  \textfont0=\eightrm \scriptfont0=\sixrm \scriptscriptfont0=\fiverm%
  \textfont1=\eighti \scriptfont1=\sixi \scriptscriptfont1=\fivei%
  \textfont2=\eightsy \scriptfont2=\sixsy \scriptscriptfont2=\fivesy%
  \textfont3=\tenex \scriptfont3=\tenex \scriptscriptfont3=\tenex%
  \def\it{\fam\itfam\eightit}\textfont\itfam=\eightit%
  \def\sl{\fam\slfam\eightsl}\textfont\slfam=\eightsl%
  \def\bf{\fam\bffam\eightbf}\textfont\bffam=\eightbf \scriptfont\bffam=\sixbf%
  \scriptscriptfont\bffam=\fivebf%
  \normalbaselineskip=9pt%
  \let\sc=\sixrm%
  \let\big=\eightbig%
  \setbox\strutbox=\hbox{\vrule height7pt depth2pt width\z@}%
  \normalbaselines\rm}

\def\tenbig#1{{\hbox{$\left#1\vbox to8.5pt{}\right.\n@space$}}}
\def\ninebig#1{{\hbox{$\textfont0=\tenrm\textfont2=\tensy
  \left#1\vbox to7.25pt{}\right.\n@space$}}}
\def\eightbig#1{{\hbox{$\textfont0=\ninerm\textfont2=\ninesy
  \left#1\vbox to6.5pt{}\right.\n@space$}}}

\def\footnote#1{\edef\@sf{\spacefactor\the\spacefactor}#1\@sf
      \insert\footins\bgroup\eightpoint
      \interlinepenalty100 \let\par=\endgraf
        \leftskip=\z@skip \rightskip=\z@skip
        \splittopskip=10pt plus 1pt minus 1pt \floatingpenalty=20000
        \smallskip\item{#1}\bgroup\strut\aftergroup\@foot\let\next}
\skip\footins=12pt plus 2pt minus 4pt 
\dimen\footins=30pc 

\newinsert\margin
\dimen\margin=\maxdimen
\def\titlefont{\seventeenpoint}
\loadtruetwelvepoint 
\loadtrueseventeenpoint
\catcode`\@=\active
\catcode`@=12  
\catcode`\"=\active

\def\eatOne#1{}
\def\ifundef#1{\expandafter\ifx%
\csname\expandafter\eatOne\string#1\endcsname\relax}
\def\notTrue{\iffalse}\def\isTrue{\iftrue}
\def\ifdef#1{{\ifundef#1%
\aftergroup\notTrue\else\aftergroup\isTrue\fi}}
\def\use#1{\ifundef#1\linemessage{Warning: \string#1 is undefined.}%
{\tt \string#1}\else#1\fi}


\global\newcount\refno \global\refno=1
\newwrite\rfile
\newlinechar=`\^^J
\def\ref#1#2{\the\refno\nref#1{#2}}
\def\nref#1#2{\xdef#1{\the\refno}%
\ifnum\refno=1\immediate\openout\rfile=\jobname.refs\fi%
\immediate\write\rfile{\noexpand\item{[\noexpand#1]\ }#2.}%
\global\advance\refno by1}
\def\lref#1#2{\the\refno\xdef#1{\the\refno}%
\ifnum\refno=1\immediate\openout\rfile=\jobname.refs\fi%
\immediate\write\rfile{\noexpand\item{[\noexpand#1]\ }#2\semi}%
\global\advance\refno by1}
\def\cref#1{\immediate\write\rfile{#1\semi}}

\def\semi{;\hfil\noexpand\break}

\def\listrefs{\vfill\eject\immediate\closeout\rfile
\centerline{{\bf References}}\bigskip\frenchspacing%
\input \jobname.refs\vfill\eject\nonfrenchspacing}

\def\inputAuxIfPresent#1{\immediate\openin1=#1
\ifeof1\message{No file \auxfileName; I'll create one.
}\else\closein1\relax\input\auxfileName\fi%
}
\def\NPB{Nucl.\ Phys.\ B}

\def\PRD{Phys.\ Rev.\ D}
\def\PLB{Phys.\ Lett.\ B}

\newif\ifWritingAuxFile
\newwrite\auxfile
\def\SetUpAuxFile{%
\xdef\auxfileName{\jobname.aux}%
\inputAuxIfPresent{\auxfileName}%
\WritingAuxFiletrue%
\immediate\openout\auxfile=\auxfileName}

\def\L{\left(}\def\R{\right)}

\def\LB{\left[}\def\RB{\right]}

\def\bye{\par\vfill\supereject%
\ifAnyCounterChanged\linemessage{
Some counters have changed.  Re-run tex to fix them up.}\fi%
\end}


\SetUpAuxFile

\def\L{\left(}
\def\R{\right)}

\def\Tr{\mathop{\rm Tr}\nolimits}

\def\pol{\varepsilon}

\def\dl^#1_#2{\delta^{#1}{}_{#2}}

\catcode`@=11  
\def\meqalign#1{\,\vcenter{\openup1\jot\m@th
   \ialign{\strut\hfil$\displaystyle{##}$ && $\displaystyle{{}##}$\hfil
             \crcr#1\crcr}}\,}
\catcode`@=12  


\baselineskip 15pt
\overfullrule 0.5pt


\def\Tr{\mathop{\rm Tr}\nolimits}

\def\pol{\varepsilon}

\def\ksl{\slashed{k}}

\def\L{\left(}\def\R{\right)}

\def\spa#1.#2{\left\langle#1\,#2\right\rangle}
\def\spb#1.#2{\left[#1\,#2\right]}
\def\lor#1.#2{\left(#1\,#2\right)}
\def\sand#1.#2.#3{%
\left\langle\smash{#1}{\vphantom1}^{-}\right|{#2}%
\left|\smash{#3}{\vphantom1}^{-}\right\rangle}
\def\sandp#1.#2.#3{%
\left\langle\smash{#1}{\vphantom1}^{-}\right|{#2}%
\left|\smash{#3}{\vphantom1}^{+}\right\rangle}
\def\sandpp#1.#2.#3{%
\left\langle\smash{#1}{\vphantom1}^{+}\right|{#2}%
\left|\smash{#3}{\vphantom1}^{+}\right\rangle}
\catcode`@=11  
\def\meqalign#1{\,\vcenter{\openup1\jot\m@th
   \ialign{\strut\hfil$\displaystyle{##}$ && $\displaystyle{{}##}$\hfil
             \crcr#1\crcr}}\,}
\catcode`@=12  

\loadfourteenpoint


%
\nopagenumbers\hsize=\hstitle\vskip1in
\overfullrule 0pt
\hfuzz 35 pt
\vbadness=10001
%
%

%

\def\lr{\leftrightarrow}

\def\Split{\mathop{\rm Split}\nolimits}

%

\baselineskip 15pt
\overfullrule 0.5pt


\noindent
hep-ph/9312333 \hfill {SLAC--PUB--6409}
\rightline{UCLA/TEP/93/51}
\rightline{December, 1993}

\leftlabelstrue
\vskip -.6 in
\Title{One-Loop N Gluon Amplitudes}
\vskip -3.7 cm
\Title{with Maximal Helicity Violation via Collinear Limits}

\centerline{Zvi Bern${}^{\flat}$ and Gordon Chalmers}
\baselineskip12truept
\centerline{\it Department of Physics}
\centerline{\it University of California, Los Angeles}
\centerline{\it Los Angeles, CA 90024}
\centerline{\tt bern@physics.ucla.edu}

\smallskip\smallskip

\baselineskip17truept
\centerline{Lance Dixon${}^{\star}$}
\baselineskip12truept
\centerline{\it Stanford Linear Accelerator Center}
\centerline{\it Stanford, CA 94309}
\centerline{\tt lance@slacvm.slac.stanford.edu}

\smallskip \centerline{and} \smallskip

\baselineskip17truept
\centerline{David A. Kosower}
\baselineskip12truept
\centerline{\it Service de Physique Th\'eorique de Saclay${}^{\dagger}$}
\centerline{\it Centre d'Etudes de Saclay}
\centerline{\it F-91191 Gif-sur-Yvette cedex, France}
\centerline{\tt kosower@amoco.saclay.cea.fr}

\vskip 0.2in\baselineskip13truept

\centerline{\bf Abstract}

\ifdraftmode
\vskip 5pt
\centerline{{\bf Draft}\hskip 10pt\TimeStamp}
\vskip 5pt
\centerline{{\bf NOT for reproduction}}
\fi

{\narrower We present a conjecture for
the $n$-gluon one-loop amplitudes with maximal helicity violation.
The conjecture emerges from the powerful requirement
that the amplitudes have the correct behavior in the collinear
limits of external momenta.
One implication is that the corresponding amplitudes where
three or more gluon legs are replaced by photons vanish for $n>4$.}

\baselineskip17pt

\centerline{\sl Submitted to Physical Review Letters}

\vfill
\vskip 0.1in
\noindent\hrule width 3.6in\hfil\break
${}^{\flat}$Research supported in part by the US Department of Energy
under grant DE-FG03-91ER40662 and in part by the
Alfred P. Sloan Foundation under grant BR-3222. \hfil\break
${}^{\star}$Research supported by the Department of
Energy under grant DE-AC03-76SF00515.\hfil\break
${}^{\dagger}$Laboratory of the {\it Direction des Sciences de la Mati\`ere\/}
of the {\it Commissariat \`a l'Energie Atomique\/} of France.\hfil\break
\Date{}

\line{}

Multi-jet processes at colliders require knowledge of matrix elements
with multiple final state partons.  At tree-level concise
formulae for maximally helicity violating amplitudes with an arbitrary
number of external legs were first conjectured by Parke and
Taylor~[\ref\ParkeTaylor{S.J. Parke and T.R. Taylor,
Phys.\ Rev.\ Lett.\ 56:2459  (1986)}],
and later proven by Berends and Giele using
recursion relations~[\ref\RecursiveBG{F.A. Berends and W.T. Giele,
Nucl.\ Phys.\ B306:759 (1988)},\ref\RecursiveK{
D.A. Kosower, Nucl.\ Phys.\ B335:23 (1990)}].

In general amplitudes in gauge theories satisfy
strong consistency conditions; they must be
unitary, and must satisfy correct limits as the momenta of external
legs become collinear~[\ParkeTaylor,\RecursiveBG,%
\ref\ColorSplitMP{M.\ Mangano and S.J.\ Parke, Nucl.\ Phys.
B299:673 (1988)}].
In this letter we discuss the example of a
one-loop amplitude which is sufficiently constrained
that we can write down a form for an arbitrary number of external legs.
The all-$n$ conjecture which we present is for maximal helicity
violation, that is with all (outgoing) legs of identical helicity,
was originally displayed in
ref.~[\ref\BDKconf{Z. Bern, L. Dixon and D. A. Kosower,
in {\it Proceedings of Strings 1993},  eds.
M.B.\ Halpern, A. Sevrin and G. Rivlis (World Scientific), hep-th/9311026}],
and has just been confirmed by recursive
techniques~[\ref\MahlonB{G.\ Mahlon,
 preprint Fermilab-Pub-93/389-T, hep-ph/9312276},\ref\Oleg{
 L. Dixon and O. Puzyrko, in preparation}].
The construction is based upon extending the known
one-loop four- and five-gluon~[\ref\FiveGluon{Z. Bern, L. Dixon and
D. A. Kosower, Phys.\ Rev. Lett.\ 70:2677 (1993)}]
amplitudes which were first obtained using string-based
methods~[\ref\StringBased{Z. Bern and D. A.\ Kosower,
Phys.\ Rev.\ Lett.\ 66:1669 (1991)
\semi
Z. Bern and D. A.\ Kosower, Nucl.\ Phys 379:451 (1992)\semi
Z. Bern and D. A.\ Kosower, in {\it Proceedings of the PASCOS-91
Symposium}, eds.\ P. Nath and S. Reucroft (World Scientific)\semi
Z. Bern and D. C. Dunbar,  Nucl.\ Phys.\ B379:562 (1992)\semi
Z. Bern, UCLA/93/TEP/5, hep-ph/9304249, proceedings of TASI 1992}].

The one-loop $n$-gluon partial
amplitude $A_{n;1}(1^+, 2^+,\ldots,n^+)$ is associated with the color
factor $N \Tr(T^{a_1} \cdots T^{a_n})$ and gives the leading
contribution to the amplitude for a large
$N$~[\ref\Color{
F. A.\ Berends and W. T.\ Giele,
Nucl.\ Phys.\ B294:700 (1987)\semi
M.\ Mangano, Nucl.\ Phys.\ B309:461 (1988)},\ColorSplitMP,%
\ref\Decoupling{Z. Bern and D. A.\ Kosower, Nucl.\ Phys.\ B362:389 (1991)}].
The subleading partial amplitudes
$A_{n;c}$, $c>1$, can be obtained from $A_{n;1}$ by summing over
various permutations [\use\Decoupling,\ref\NeqFour{
 Z. Bern, L. Dixon, D. Dunbar and D.A. Kosower, preprint
 SLAC--PUB--6415}].
The structure of $A_{n;1}$
is particularly simple,
making it an ideal candidate for finding an all $n$ expression.
The all-plus helicity structure is cyclicly symmetric;
and no logarithms or other functions containing branch cuts can appear.
This can be seen by considering the cutting rules:
the cut in a given channel is given by
a phase space integral of the product of the two tree amplitudes
obtained from cutting.  One of these tree amplitudes will vanish for all
assignments of helicities on the cut internal legs since
$A_{n}^{\rm tree} (1^\pm,2^+,3^+,\ldots,n^+)=0$, so that all cuts
vanish. Similar reasoning shows that
the all plus helicity loop amplitude does
not contain multi-particle poles.
The only singularities are those where two (color-adjacent) momenta
become collinear.

Another simplifying feature of the all-plus amplitude is the equality,
up to a sign due to statistics, of the contributions
of internal gluons, complex scalars and Weyl fermions.
This is a
consequence of the supersymmetry Ward identity~[\ref\Susy{
M.T.\ Grisaru, H.N.\ Pendleton and P.\ van Nieuwenhuizen,
Phys. Rev. {D15}:996 (1977)\semi
M.T. Grisaru and H.N. Pendleton, Nucl.\ Phys.\ B124:81 (1977)\semi
S.J. Parke and T. Taylor, Phys.\ Lett.\ B157:81 (1985)\semi
Z. Kunszt, Nucl.\ Phys.\ B271:333 (1986)}]
$A^{\rm susy}(1^\pm,2^+,\ldots,n^+) = 0$ for $N=1$ and $N=2$ theories.
For Weyl fermions and complex scalars transforming under the fundamental
rather than the adjoint representation (in a vector-like theory), the color
factor is smaller by a factor of $N$, and {\it no\/} subleading color
factors appear.

At one loop the collinear limits of color-ordered one-loop QCD amplitudes
are expected to have the form
$$
\eqalign{
A_{n;1}^{\rm loop} \mathop{\longrightarrow}^{a \parallel b}
\sum_{\lambda=\pm}  \biggl(
  \Split^{\rm tree}_{-\lambda}(a^{\lambda_a},b^{\lambda_b})\,
&
      A_{n-1;1}^{\rm loop}(\ldots(a+b)^\lambda\ldots)
\cr
&  +\Split^{\rm loop}_{-\lambda}(a^{\lambda_a},b^{\lambda_b})\,
      A_{n-1}^{\rm tree}(\ldots(a+b)^\lambda\ldots) \biggr) \;,
\cr}
\eqn\loopsplit
$$
in the limit where the momenta $k_a \rightarrow z k_P$ and $k_b \rightarrow
(1-z) k_P$ with $k_P = k_a + k_b$. Here $\lambda$ is the helicity
of the intermediate state with momentum $k_P$.
This is analogous to the form of tree-level collinear limits
[\use\ParkeTaylor,\use\RecursiveBG,\use\ColorSplitMP,%
\ref\ManganoReview{M. Mangano and S.J. Parke,
 Phys.\ Rep.\ 200:301 (1991)}].
The explicit form of the one-loop splitting
functions may be extracted from the known
four-~[\ref\Long{Z. Bern and D. A. Kosower, Nucl.\ Phys.\ B379:451 (1992)}]
and five-point [\use\FiveGluon] gluon amplitudes.
All known one-loop amplitudes~[\FiveGluon,\use\NeqFour]
satisfy eq.~(\use\loopsplit), though there is as yet
no proof of its correctness for larger $n$.
Because of the supersymmetry Ward identitity
relating the gluon and fermion contribution to the scalar one,
it suffices for our present purposes to prove it
for the case of scalars in the loop.

The one-loop all-plus helicity amplitudes have a simple collinear
structure because the loop splitting function
Split$^{\rm loop}_{-\lambda}$ does not enter;
it multiplies a tree amplitude which vanishes.
The tree splitting functions that enter
are~[\ParkeTaylor,\RecursiveBG,\ColorSplitMP]
$$\eqalign{
\Split^{\rm tree}_{+}(a^{+},b^{+})\ &=\ 0, \hskip 1.2 cm
\Split^{\rm tree}_{-}(a^{+},b^{+})
            = 1/ (\sqrt{z (1-z)}\spa{a}.b) , \cr}
\eqn\gggtree
$$
where we follow the notation of ref.~[\ManganoReview] for
the spinor inner products $\spa{a}.b$ and $\spb{a}.b$ which
are equal to $\sqrt{s_{ab}}$ up to a phase.
In general, the non-vanishing splitting functions diverge as
$1/\sqrt{s_{ab}}$ in the collinear limit $s_{ab}=(k_a+k_b)^2\rightarrow0$.

We now outline a proof of the universality of the scalar-loop
contributions to the collinear splitting
functions.  We divide the
diagrams into several sets,  depending upon the
topology of the two external collinear legs which, without loss of
generality, we label~1 and 2.
In a color-ordered diagram, only adjacent legs
can have collinear singularities.
It turns out that $\Split^{\rm tree}$ arises from the
diagrams in fig.~1, $\Split^{\rm loop}$ from the diagrams in fig.~2 and
diagrams without explicit poles in $s_{12}$, such as those of fig.~3,
do not contribute to the splitting functions.

We begin with the diagrams in fig.~1.
The only Feynman diagrams which can contribute to the tree
splitting function are those
containing explicit poles in $s_{12}$, as depicted
in fig.~1; trees containing legs 1 and 2 but lacking this explicit
pole will not contribute.  The analysis is identical to the tree-level
analysis and gives a similar result, yielding
the first term in eq.~(\use\loopsplit) containing the tree splitting
function.

The diagrams in fig.~2 also contain
explicit collinear poles and  give rise to the $\Split^{\rm loop}$
function. There are three groups of diagrams in this category
depicted in figs.~2a--c.
Evaluating and summing over the three types of diagrams in the collinear
limit yields
$$
{1 \over 16\pi^2} {1\over 6} ~(k_1-k_2)^{\mu}
\eta_{\mu\nu}
A^{\rm tree}_{n-1}(1+2,\ldots,n)^{\nu} \left({\sqrt{2} \over
s_{12}}\right) \left[ \pol_1 \cdot \pol_2 - {\pol_1
\cdot k_2 \pol_2 \cdot k_1 \over k_1 \cdot k_2} \right],
\anoneqn
$$
where $\pol_i$ are gluon polarization vectors.
This will give the entire contribution to the loop splitting
functions for internal scalars.
Converting to a helicity basis
[\ref\Helicity{
F.\ A.\ Berends, R.\ Kleiss, P.\ De Causmaecker, R.\ Gastmans and T.\ T.\ Wu,
        Phys.\ Lett.\ 103B:124 (1981)\semi
P.\ De Causmaeker, R.\ Gastmans,  W.\ Troost and  T.\ T.\ Wu,
Nucl. Phys. B206:53 (1982)\semi
R.\ Kleiss and W.\ J.\ Stirling,
   Nucl.\ Phys.\ B262:235 (1985)\semi
   J.\ F.\ Gunion and Z.\ Kunszt, Phys.\ Lett.\ 161B:333 (1985)\semi
Z. Xu, D.-H.\ Zhang and L. Chang, Nucl.\ Phys.\ B291:392 (1987)}]
in a manner similar to that
used at tree-level in ref.~[\RecursiveBG], one finds
$$\hskip-20pt\eqalign{
\Split^{{\rm loop}\ [0]}_{+}(a^{+},b^{+})
\ &=\ -\sqrt{z(1-z)} {\spb{a}.b
  /( 48\pi^2 {\spa{a}.b}^2}) \,,  \cr
\Split^{{\rm loop}\ [0]}_{-}(a^{+},b^{+})
\ &=\ \sqrt{z(1-z)}/({48\pi^2\spa{a}.b})\, , \cr}
\anoneqn
$$
and $\Split^{\rm loop}_{-\lambda}(a^\pm,b^\mp)$ vanishes.

The remaining diagrams do not
have the required collinear pole arising from a tree propagator; it
would have to emerge from the loop integral.
One possibility is that one collinear leg
is directly connected to the loop a via a three vertex while the other
collinear leg is part of a tree or a four-vertex sewn onto the loop.
These diagrams cannot have any collinear poles in $s_{12}$
because the loop integral does not contain this kinematic invariant
except as a sum with other kinematic invariants.

The next possibility, depicted in fig.~3a, is that both legs in the
collinear pair are attached to a scalar loop by three-point vertices
and are part of a loop with four or more legs.
Since the splitting functions diverge as $s_{12} \rightarrow 0$,
contributions come from regions where the three propagators
$1/(l-k_2)^2$, $1/l^2$, and $1/(l+k_1)^2$ depicted in fig.~3a
blow up.
The leading singularities
come from the region $l\approx \alpha k_1 + \beta k_2$ where
$\alpha$ and $\beta$ are arbitrary
constants. Near the special points $(\alpha,\beta)=(-1,0)$ and
$(0,1)$ a fourth propagator blows up requiring a separate analysis,
which will lead to the same conclusion as the generic case.
In the generic case, in the region $l\approx \alpha k_1 + \beta k_2$
the calculation reduces to a triangle integral.
An analysis of the integral~[\ref\FutureColl{
 Z. Bern and G.\ Chalmers, in progress}]
shows that there are no contributions to the splitting functions
from fig.~3a or b.
For gluons or fermions circulating in the loop (for a generic helicity
amplitude), loop-momentum-independent
terms in the vertices of the diagrams in fig.~3 invalidate the
above analysis.

The starting point in constructing our $n$-point expression is the known
five-point one-loop helicity amplitude~[\use\FiveGluon],
$$
\eqalign{
A_{5;1}(1^+,2^+,3^+,4^+,5^+)\  &=\ {iN_p\over 192\pi^2}\,
  {  s_{12}s_{23} + s_{23}s_{34} + s_{34}s_{45} + s_{45}s_{51} +
     s_{51}s_{12}\ +\ \pol(1,2,3,4)
   \over \spa1.2 \spa2.3 \spa3.4 \spa4.5 \spa5.1 }\ ,  \cr }
\anoneqn
$$
where
$\pol(i,j,m,n) = 4i\varepsilon_{\mu\nu\rho\sigma}
        k_i^\mu k_j^\nu k_m^\rho k_n^\sigma
    \ =\ \spb{i}.{j}\spa{j}.{m}\spb{m}.{n}\spa{n}.{i}
       - \spa{i}.{j}\spb{j}.{m}\spa{m}.{n}\spb{n}.{i}
$, and $N_p$ is the number of color-weighted bosonic states
minus fermionic states circulating in the loop;
for QCD with $n_f$ quarks, $N_p = 2(1-n_f/N)$ with $N=3$.

\def\ksl{\slashed{k}}
Using eqs.~(\use\loopsplit) and (\use\gggtree) and
$A^{\rm tree}_n (1^\pm, 2^+, \cdots, n^+) = 0$,
we can construct higher point amplitudes
by writing down general forms
with only two particle-poles,
and requiring that they have the correct collinear limits.
Generalizing to all $n$ we have
$$\eqalign{
 A_{n;1}(1^+,2^+,\ldots,n^+)\ =\ {i N_p \over 192\pi^2}\,
{E_n + O_n \over \spa1.2 \spa2.3 \cdots \spa{n}.1 }\ ,
}\eqn\allnplus
$$
where
$$
  O_n\ =\ \sum_{1\leq i_1 < i_2 < i_3 < i_4 \leq n-1}
            \pol(i_1,i_2,i_3,i_4)
    \ =\ -\sum_{1\leq i_1 < i_2 < i_3 < i_4 \leq n}
           {\rm tr}(\ksl_{i_1}\ksl_{i_2}\ksl_{i_3}\ksl_{i_4}\gamma^5)
  \; .
\eqn\oddansatz
$$
To describe $E_n$ define
$t^{[p]}_i = (k_i+k_{i+1}+\cdots+k_{i+p-1})^2$
(all indices mod $n$); note that
$t^{[2]}_i = s_{i,i+1}$ and
$t^{[1]}_{i}=0$.  Then
$$
E_{n}\ =\ \sum_{p=2}^{\lfloor (n-1)/2 \rfloor}
\sum_{i=1}^{n}  \L t^{[p]}_{i}  t^{[p]}_{i+1}
                   - t^{[p-1]}_{i+1}  t^{[p+1]}_{i}\R
\ +\ \LB {1\over2}\sum_{i=1}^{n}  \L t^{[m]}_{i}  t^{[m]}_{i+1}
                   - t^{[m-1]}_{i+1}  t^{[m+1]}_{i}\R \RB_{m=n/2,\
  n\ {\rm even}}
\eqn\evenansatz
$$
or equivalently
$$
E_n  = -\sum_{1\leq i_1 < i_2 < i_3 < i_4 \leq n}
           {\rm tr}(\ksl_{i_1}\ksl_{i_2}\ksl_{i_3}\ksl_{i_4})
  \; .
\eqn\altevenansatz$$
The two terms $O_n$ and $E_n$ can be combined into a single
trace, a form which agrees with ref.~[\use\MahlonB],
but for the purposes of discussing symmetry properties, it
is more convenient to keep them separate.

The $O_n$ term~(\oddansatz) is not manifestly cyclicly symmetric;
however, the difference between $O_n$ and its cyclic permutation
vanishes using momentum conservation.
To verify that in the limit that two legs become collinear
it reduces to the corresponding
$(n-1)$-point term $O_{n-1}$,
it suffices to check the limit $1\ ||\ 2$.
Terms of the form $\pol(1,2,j_3,j_4)$ clearly vanish.
The remaining terms containing $1$ and $2$ may be paired as
$
  \pol(1,i_2,i_3,i_4) + \pol(2,i_2,i_3,i_4) 
  \ =\
\pol(P,i_2,i_3,i_4),
$
where $k_P=k_1+k_2$.  Adding these terms to the terms containing
neither $1$ nor $2$, and relabeling
$\{P,3,4,\ldots,n\} \to \{1,2,3,\ldots,n-1\}$, we see that
$O_n \to O_{n-1}$ in the limit $1\ ||\ 2$, as required.
The cyclic symmetry of the $E_n$ term~(\evenansatz) is manifest.
The collinear limit of the equivalent form~(\use\altevenansatz)
follows the same argument as for the $O_n$ terms.

Assuming that the denominator of the all-plus amplitude is given by
$\spa1.2\cdots\spa{n}.1$, one can prove that the functions $E_n$ and
$O_n$ are uniquely determined by the collinear limits for all $n>5$.
(The collinear limit of $O_5$ is special because $\pol(1,2,3,4)$
vanishes in all collinear limits.)  Presumably one should be able to
give a proof of the same statement relaxing the denominator assumption.

\font\eightsy=cmsy8
\def\soft#1#2#3{{\cal S}_{#2}(#1,#3)}

In massless QED, through use of recursion
relations~[\use\RecursiveBG,\RecursiveK], Mahlon has demonstrated that
the one-loop $n$-photon helicity amplitudes $A_n(\gamma_1^\pm,
\gamma_2^+, \cdots,
\gamma_n^+)$ vanish for $n>4$ [\ref\MahlonA{G.\ Mahlon,
 preprint Fermilab-Pub-93/327-T, hep-ph/9311213}].
One can generalize Mahlon's results in the all-plus case
to `mixed' photon-gluon amplitudes using
the expression~(\allnplus) and converting some of the gluons into
photons.
Amplitudes with $r$ external photons and $(n-r)$ gluons have a color
decomposition similar to that of the pure-gluon amplitudes,
except that charge matrices are set to unity for the photon legs.
The coefficients of these color factors, $A_{n;1}^{r\gamma}$,
are given by appropriate cyclic sums over the pure-gluon partial
amplitudes, retaining only the contributions from particles in the
fundamental representation in the loop;
e.g., for a single quark with electric charge $Q$, replace
$N_p \to N_p^{\rm fund} = -2/N$, and the overall coupling factor
$g^n \to g^{n-r} (e Q \sqrt2)^r$.
Defining the short-hand
$
\soft{i}{n}{j}\ =\
{{\spa{i}.{j}/(\spa{i}.{n}\spa{n}.{j})}}
$,
performing the cyclic sums, and making repeated use of spinor identities
we can write down simple forms for the all-plus
partial amplitude with one or two external photons (legs $n\ldots n-r+1$),
and any number of gluons,
$$
 A_{n;1}^{r\gamma}\ =\ {i N_p^{\rm fund} \over 192\pi^2}\,
{O^{r\gamma}_n+E^{r\gamma}_n
 \over\spa1.2\spa2.3\cdots\spa{n-r-1,}.{n-r}\spa{n-r,}.1}\,,
\anoneqn
$$
with
$$\eqalign{
O^{1\gamma}_n &= -2
\sum_{1\leq i_1 < i_2 < i_3 < i_4 \leq n-1}
         \pol(i_1,i_2,i_3,i_4)\,\LB \soft{i_1}n{i_2}+\soft{i_3}n{i_4}\RB\,,\cr
E^{1\gamma}_n &= 2\ \sum_{1\leq i_1 < i_2 < i_3 \leq n-1}
\LB \soft{i_1}{n}{i_2} \, s_{i_1i_2} s_{i_3n}
  + \soft{i_2}{n}{i_3} \, s_{i_2i_3} s_{i_1n}
  + \soft{i_3}{n}{i_1} \, s_{i_3i_1} s_{i_2n} \RB\,,\cr
O^{2\gamma}_n &= 4
\sum_{1\leq i_1 < i_2 < i_3 \leq n-2}
            \pol(i_1,i_2,i_3,n-1)\,\LB \soft{i_1}n{i_2}\soft{i_2}{n-1}{i_3}
                             -\soft{i_1}{n-1}{i_2}\soft{i_2}{n}{i_3}\RB\,,\cr
E^{2\gamma}_n &= 4\ \sum_{1\leq i_1 < i_2 \leq n-2}
\LB \soft{i_1}{n-1}{i_2}\soft{i_1}{n}{i_2}
{\rm tr}(\ksl_{i_1}\ksl_{n}\ksl_{i_2}\ksl_{n-1})
- s_{i_1 i_2}{\spb{n-1}.n\over\spa{n-1}.n}\RB\,.\cr
}\anoneqn
$$

For three or more external photons, an even more striking result
emerges: the amplitude vanishes,
$$
A^{\rm loop}_{n>4}(\gamma_1^+, \gamma_2^+, \gamma_3^+, g_4^+, \ldots, g_n^+)
 =0.
\anoneqn
$$
Since amplitudes with even more photon legs are obtained by further sums
over permutations of legs, the all-plus helicity amplitudes with
three or more photon legs vanish
(for $n>4$) in agreement with the expectation from the collinear limits.

In order to extend these methods to other helicity amplitudes
one would first need a general proof of the collinear limits for particles
circulating in the loop other than scalars~[\use\FutureColl] (which
sufficed for the all-plus case because of the supersymmetry identities).
The loop splitting functions appearing in equation~(\loopsplit) can
already be extracted from five-parton amplitudes~[\use\FiveGluon,\use\NeqFour].
We expect that collinear limits will be a useful tool in constructing
one-loop helicity amplitudes besides those presented here.

\listrefs

\vfill\break

\centerline{\bf Figure Captions.}

\vskip .5 cm

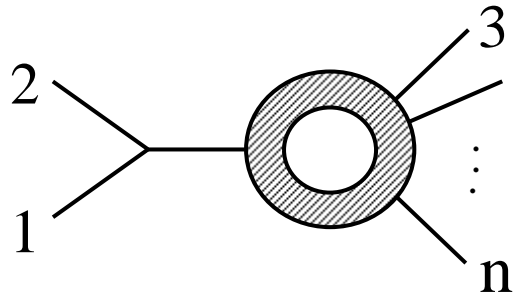
\item{\bf Fig.\ 1:} Diagrams that contribute to the tree splitting functions.

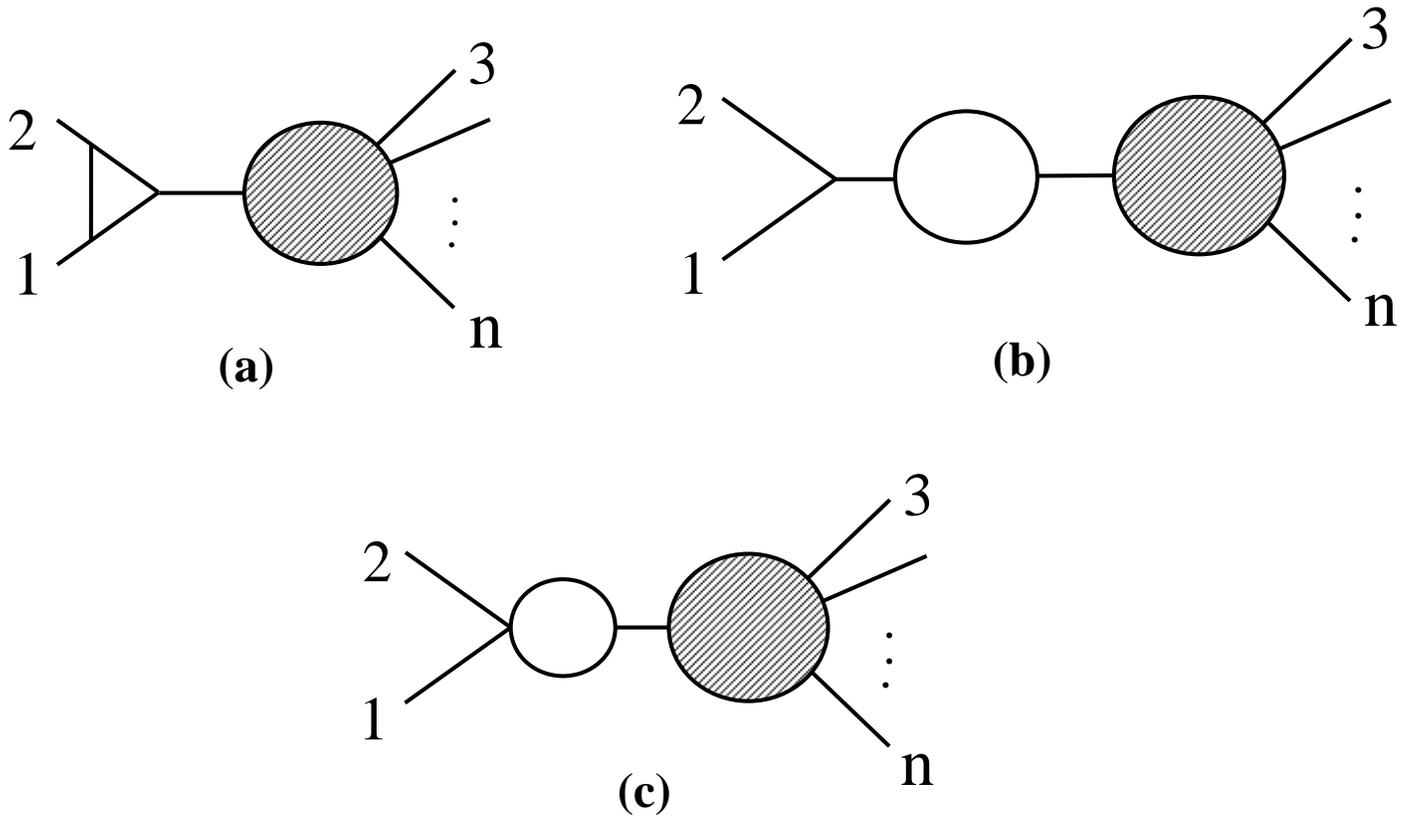
\item{\bf Fig.\ 2:} Diagrams that contribute to the loop splitting functions.

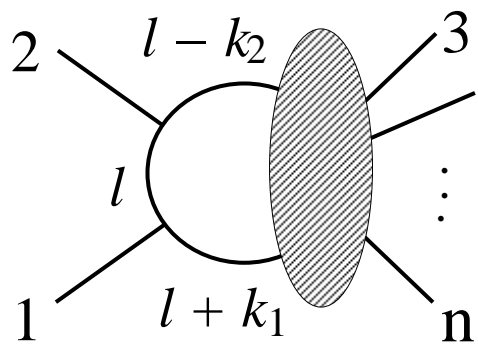
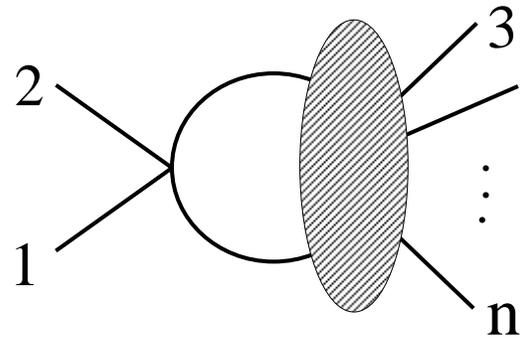
\item{\bf Fig.\ 3:} Two of the remaining diagram types which have no
collinear poles for scalars in the loop.

\bye